\documentstyle[preprint,aps]{revtex} 
\begin{document}
\draft

\title{Stokes shift in quantum wells: trapping versus thermalization}

\author{A. Polimeni, A. Patan\`e, M. Grassi Alessi, and M. Capizzi}
\address{Istituto Nazionale di Fisica della Materia - Dipartimento di Fisica, 
Universit\'a di Roma "La Sapienza", P.le A. Moro 2, I-00185 Roma, Italy}

\author{F. Martelli}
\address{Fondazione Ugo Bordoni, via B. Castiglione 59, I-00142 Roma, Italy}

\author{A. Bosacchi and S. Franchi}
\address{Consiglio Nazionale delle Ricerche, Istituto Materiali Speciali per
l'Elettronica e Magnetismo, via Chiavari 18/A, I-43100 Parma}

\date{\today}
\maketitle

\begin{abstract}
Low temperature photoluminescence and photoluminescence excitation measurements
have been performed in a set of InGaAs/GaAs samples with different indium
molar fraction, well width, growth conditions and post-growth treatment.
This has allowed to change in a controlled way the degree and source of  
disorder in the samples, thus resulting in an excitonic absorption linewidth
varying between 1 and 18 meV, and an ensueing Stokes shift changing between
zero and 8 meV. The conditions of validity of two different models relating the 
Stokes shift to the linewidth broadening have been established in terms of 
different regimes of disorder and temperature. A continuous transition between 
those regimes has been demonstrated.
\end{abstract} 

The full width at half maximum (FWHM) of the heavy-hole free exciton (HHFE)
recombination and the Stokes shift {\it SS}, i.e., the difference between the
exciton peak energy as determined by photoluminescence (PL) and absorption
measurements, are commonly considered indicative of the sample quality. 
However, the origin of the Stokes shift is still argument of debate, although 
it has been the subject of a number of experimental and theoretical
works.\cite{Bastard,Delalande,Yang,Wilkinson,Ping,Gurioli,Humlicek} 
In the pioneering works in GaAs/Al$_x$Ga$_{1-x}$As single quantum wells
(QWs),\cite{Bastard,Delalande} the {\it SS} was 
attributed to trapping of excitons at interface defects with a contribution 
to the interface potential described by a gaussian function. Assuming a lack 
of exciton thermalization, i.e., a recombination of excitons at the defect 
sites where they are first trapped, the {\it SS} is determined by the 
distribution of the binding energy of excitons to the defects. The vanishing 
of {\it SS} for increasing temperature has been attributed to an increased 
detrapping of excitons. In later works, the main cause of exciton trapping to a
disordered potential has been identified in the interface roughness produced by
an inhomogeneous thickness of the QWs.\cite{Yang,Wilkinson} In other
theoretical models, a major emphasis has been given, instead, to thermal 
effects on carrier migration,\cite{Ping} the carrier 
thermalization\cite{Gurioli} or the thermodynamical balance between excitons 
and radiation.\cite{Humlicek}  We will briefly resume in the 
following those models which lead to an analytical relationship between 
{\it SS} and the value {\it W} of FWHM of the absorption line and thus allow a 
comparison with experimental data.

In the model of Fang Yang {\it et al},\cite{Yang} the probability distribution
of the exciton energy in the QW-plane (x,y) is assumed to be a function G(x,y),
which determines the shape of the absorption band. Fluctuations in the QW 
thickness, with a gaussian distribution due to the diffusion process during 
the QW growth,\cite{Wilkinson} are assumed to be the only source of randomness 
in the exciton energy distribution. If this topographical disorder extends 
over a scale much greater than the carrier diffusion length, carriers relax 
into local minima before recombining as excitons. Therefore, the PL reflects 
the distribution of local minima in the exciton-energy distribution and a 
{\it SS} ensues, with

\begin{equation}             
SS = 0.55\ W\, .
\end{equation}

A linear relationship is found, indeed, between {\it SS} and {\it W} as 
obtained in several 
III-V and II-VI compounds, with a proportionality factor equal to 0.6. It is 
worth noticing that the values of $W$ range from a few meV to about 100 meV, 
suggestive of a medium to high degree of disorder. A similar linear behavior,
although affected by a sizeable scattering in the data, has been also obtained 
in the case of symmetrically strained (GaIn)As/Ga(PAs) superlattices for 
{\it W} varying between 6 and 10 meV.\cite{Lutgen}  

If thermal effects in the exciton diffusion are added to a gaussian disorder,
only numerical estimates can be carried out.\cite{Ping} A linear relationship 
between {\it SS} and the PL linewidth is found to hold in the low temperature 
limit, with a slope smaller (0.3) than that found in absence of thermal 
effects. No comparison with experimental results is provided by the authors.

In the model of Gurioli and coworkers,\cite{Gurioli} the emphasis is further 
shifted toward the role of thermal effects. Excitons are assumed to be in 
thermal quasiequilibrium, with a distribution characterized by an effective 
carrier temperature $T_c$ greater or equal to that of the lattice. The Stokes 
shift is then related to the degree of carrier thermalization, as measured 
by the ratio between the carrier thermal energy k$_BT_c$ and the inhomogeneous 
broadening of the exciton-band. Therefore, {\it SS} becomes vanishingly 
small whenever that ratio is much greater than unity (namely, when a high
temperature allows to establish a thermal equilibrium). Otherwise, the high 
energy side of the exciton distribution function is depressed in PL because of 
population effects, and a sizeable {\it SS} results. Provided that the disorder
responsible for the exciton broadening is described by a gaussian function, 
the model results into

\begin{equation}                                  
SS = 0.18\ W^2/k_BT_c\, ,
\end{equation}             

\noindent
in quite good agreement with experimental data in AlGaAs/GaAs QWs where {\it W}
is of the order of a few meV or less.
Therefore, both models described above account for the origin of the Stokes 
shift if applied to systems which largely differ in the degree (and nature) of 
disorder. 

Recently, our group has studied the effects of structural disorder on
the photoluminescence of strained InGaAs/GaAs quantum wells grown by molecular 
beam epitaxy (MBE)\cite{Patane,Martelli} and differing in the indium molar 
fraction {\it x} and/or well width {\it L}. A detailed analysis of the PL 
lineshape for increasing {\it x} and {\it L} has evidentiated the role and 
weight of alloy disorder and well-width fluctuations $\relbar$ namely, of 
the main mechanisms responsible for the spectral broadening of HHFE 
recombination.\cite{Patane} Potential fluctuations give also rise to 
localization of excitons, most likely at the interface.\cite{Martelli} 
In the present work, we will compare the two analytical models on the origin of 
{\it SS} previously resumed with experimental results in those strained 
InGaAs/GaAs
heterostructures. In this case, the HHFE peak energy and absorption linewidth 
are obtained from PL excitation (PLE) measures because of the low absorbance of
quantum wells.\cite{nota1} By tuning the amount and source of disorder in the 
InGaAs/GaAs QWs, we will show that both the topographical 
model\cite{Yang,Wilkinson} and the quasiequilibrium model\cite{Gurioli} hold in
this system. Those models, indeed, well describe the experimental data in 
the two opposite limits of large or small $W$ (disorder), respectively, with 
a crossover between the two regimes for $W\sim 6$ meV. Moreover, for increasing
lattice temperatures data turn out to be well described by the quasiequilibrium
model also in cases where the topographical model holds at low temperature.

The samples used for this work are In$_x$Ga$_{1-x}$As/GaAs single QWs grown on 
GaAs(100) substrates, with {\it x} = 0.09, 0.10, 0.19, 0.80, and 1.0 and 
different well widths (see Table I). Only single QWs have been considered in 
order to 
eliminate possible inter-well fluctuations. The {\it x} = 0.09, 0.19 and 
{\it x} = 1.0 samples, the same used in Ref. 9, have been grown by MBE in a 
Varian GenII machine at 520 $^\circ$C; further details on the {\it x} = 1.0 
InAs/GaAs sample-growth are given elsewhere.\cite{Polimeni94} Samples with 
{\it x} = 0.10 ($x$ = 0.80) were grown in a different MBE machine at 
360 $^\circ$C (460 $^\circ$C), in order to reduce both In
segregation\cite{Bosacchi} and In surface migration. 
The same applies to samples irradiated with deuterium by a Kauffman source, as 
discussed in the following. 
The PL measurements have been performed by exciting the samples with an Ar$^+$
laser. A Ti:Sapphire laser has been used as light source in the PLE
measurements. In both cases the luminescence was dispersed through a
65-cm double-grating monochromator and detected by standard photon counting
tecniques. The samples were held at temperatures ranging between 5 K and 70 K.

In Figure 1, we report typical PL and PLE spectra showing the HHFE transition
for different indium concentrations (the values of $W$ and of {\it SS}
for all samples investigated are given in Table I). For ease of comparison, 
the PL and PLE spectra have been reported by taking for 
each sample the zero of the energy at the PLE peak. The Stokes shift and 
the recombination and absorption broadenings are clearly correlated, both 
increasing from the top to the bottom of the figure. The shoulder on the low 
energy side in the spectrum of the sample shown in Fig. 1(a) is due to an 
exciton bound to interfacial disorder.\cite{Martelli} The peak at about 10 meV 
above the HHFE in the sample reported in Fig. 1(b) is due to the 
partially-forbidden transition e1-hh3, usually observed in wide 
QWs.\cite{Lambkin} In Fig. 1(c), the low- and high-energy bands in the PL 
spectrum are due to the recombination of the carriers in the quantum dots 
formed during the growth and in the flat InAs QW, 
respectively.\cite{Polimeni96} In the InAs/GaAs QWs, both PL and PLE 
measures have been performed on deuterated samples, where a giant increase in 
the luminescence efficiency makes measures easier.\cite{Polimeni94}
It is worth noting that the diffusion of deuterium, besides increasing the PL 
signal, produces an increased degree of disorder, as evidentiated by a 
sizeable brodeaning of the luminescence and absorption recombination bands (not
shown here).

In Figure 2, we plot the values of {\it SS} as a function of {\it W} as
measured in the PLE spectra for the different samples. The data corresponding
to high values of $SS$, namely, where the disorder dominates the PL spectra, 
fall on a straight line as expected from the model introduced in Refs. 3 and 4.
Some scattering in the data appears instead for low values of {\it W}, 
as reported also in Ref. 8 for the same range of $W$s. The full line 
reported in the figure for $W \ge 6$ meV is a best fit of a linear function to 
the data ($SS$ = - 0.6 + 0.48\ $W$).
The linear behavior, as well as the slope, are in good agreement with the 
relation found in Ref. 3, although the fit accounts for null values of {\it SS}
at finite values of $W$, a feature not predicted by the model.\cite{nota2}  

Let us now give a closer inspection to the region of small {\it W} 
($\le$ 6.0 meV, or $SS\le$ 2.0 meV), where the linear 
extrapolation from the high disorder regime (dashed line in Fig. 2) 
gives a poor fit to the data and the validity 
of the linear model has been questioned.\cite{Gurioli} In Figure 3, 
we compare the experimental results with the Eq. (2) of the thermalization 
model, where $T_c$ has been estimated from the high-energy tail of the PL 
spectra. The thermalization model well accounts for the experimental results in
this region of low $W$ values. 
A quadratic relationship between {\it SS} and $W$ does not hold, instead, for 
the whole set of data, as shown in the inset where the model clearly deviates
from the experimental data for $W \ge 6$ meV.

As from Figures 2 and 3, our data agree better with one model or the other
according to the values of {\it SS} and FWHM, namely, to the sample 
``quality''. This is a consequence of the different physical mechanisms 
introduced in the two models to explain the red shift with respect to the 
absorption peak of the HHFE recombination line. 
If one assumes that the thermalization is the predominant mechanism, all the 
energy levels contributing to the absorption are in thermal equilibrium and 
the carriers diffuse to the lowest energy state available before they 
recombine. As a matter of fact, the experimental data are explained by the 
thermalization model in the samples where the thermal energy is about equal to 
the FWHM (e. g., {\it x} = 0.19 and {\it x} = 0.09 for small {\it L}, see 
Table I) and a stronger communication between the broadened excitonic levels 
is expected. On the other hand, the carrier diffusion becomes less effective 
whenever the sample inhomogeneities give rise to local energy minima and to 
potential fluctuations large with respect to the carrier thermal energy. 
In this case, if hopping processes between local
minima are ignored (see discussion in Ref. 17), the different energy levels
do not interact with each other and the PL is ruled by the distribution of
heights of local minima, as stated by Yang and coworkers. 
In the InAs/GaAs QWs, the fluctuations of {\it L} (at least 
one ML) is quite large with respect to its mean value (2 ML or less). Moreover, 
quantum dots self-aggregate at the interface\cite{Polimeni96} thus 
giving rise to a further source of disorder. In this case, the thermal energy 
turns out to be quite lower than {\it W} thus
satisfying the basic assumption of deep local minima associated with a strong 
disorder. 

Let us now comment the limit of validity of the topographical model.
In Ref. 9, it has been shown that alloy and interface disorder account for the
PL linewidth. The lateral extent of the potential fluctuations at the interface
has been estimated in about 2 nm, one order of magnitude smaller than the 
exciton Bohr radius ($\sim$ 20 nm). This results in a single, usually narrow 
HHFE recombination line. A very similar picture has been given recently in a 
theoretical paper.\cite{Efros} Broader HHFE recombination lines can be expected
when growing the samples at a lower temperature, because a lower In surface
migration increases the interface- and alloy-disorder. The samples grown at
360 
$^\circ$C (x = 0.10) have, indeed, a larger linewidth with respect to 
samples\cite{Patane} grown at 520 $^\circ$C, especially for 
wide $L$ (this last feature suggests that the low temperature
growth affects the alloy disorder more than the interface roughness). Moreover, 
also a deuterium diffusion into the samples results in an increased disorder,
most likely on a short-scale. The disorder accounting for the FWHM in 
InGaAs/GaAs QWs is therefore always on a scale length smaller than
the exciton dimension.
In Refs. 3 and 4, it is stated that the fluctuations in the potential are 
dominated by variations of the effective width over length scale greater than 
the exciton size, whereas variations of the well width over smaller dimensions
are smeared out by the finite extension of the exciton wavefunction. However, 
this does not prevent that model from holding in the present case.
In fact, although individual excitons feel potential fluctuations
averaged over a scale length smaller than their radius, the mean
potential value found on the exciton radius fluctuates over the whole
sample.\cite{Patane} Flat regions on the scale of the exciton size, which 
would give rise to multiple PL peaks,\cite{Kopf} not present in our samples, 
are therefore not a mandatory feature of the topographical model.
 
A further support to the picture of a continuous transition between the
quasiequilibrium and the topographical disorder model is provided by the 
dependence on temperature of the
Stokes shift reported in Fig. 4(a) for the InAs/GaAs sample with $L$ = 1.2 ML. 
At 5 K, this sample has a PLE linewidth equal to 13 meV (see Table I) and 
belongs, therefore, to the samples where the topographical model better works.
However, when the temperature increases from 5 K to 70 K, $SS$ decreases from 
6 to 4.1 meV, a behavior which cannot be explained by the topographical 
model. On the other hand, an increasing value of the ratio k$_BT_c/W$ between 
the thermal energy and the disorder may convert a regime where the local 
disorder dominates the spectra into a regime where the carrier thermalization 
is predominant. This is, indeed, the case of the present sample, as shown 
in Fig. 4(b) where the difference $\Delta SS$ between the measured value of 
the Stokes shift and that predicted by the quasiequilibrium model, renormalized
with respect to the experimental value, is reported vs k$_BT_c/W$. As expected, 
$\Delta SS/SS$ decreases for increasing k$_BT_c/W$.
             
In conclusion, photoluminescence and photoluminescence excitation measurements 
in InGaAs/GaAs quantum wells whose morphology has been gradually varied by 
changing the indium concentration, the well width, the growth conditions 
and/or post-growth treatments, have shown that different functional 
relationships 
between the {\it SS} and the FWHM hold for different degree of interplay 
between thermalization and disorder. In samples with low In-concentration, the 
thermalization process between the inhomogeneously broadened absorption band 
dominates. When the FWHM is large enough to make thermalization unlikely, the 
topographical model proposed by Yang and coworkers applies.
Measurements as a function of temperature have also confirmed that those two
regimes can be continuously converted one into the other. 


\begin{table} 

\caption{Values of the {\it SS} and of the linewidth of PLE for different 
indium concentration $x$, well width {\it L} and growth temperature $T_G$}

\begin{tabular}{ccccc}

    $L$       &     $x$     &      $T_G$     &     $SS$         &      $W$   \\
     (nm)     &             &   ($^\circ$C)  &     (meV)        &     (meV)  \\
\tableline
     5.0      &     0.09    &       520      &     0.0          &      1.2   \\
     5.0      &     0.10    &       360      &     0.8          &      2.8   \\
    10.0      &     0.10    &       360      &     0.7          &      3.6   \\
    20.0      &     0.10    &       360      &     1.8          &      6.0   \\
     1.5      &     0.19    &       520      &     0.3          &      1.4   \\
     2.0      &     0.19    &       520      &     0.7          &      1.6   \\
     3.0      &     0.19    &       520      &     1.4          &      3.8   \\
     4.0      &     0.19    &       520      &     1.6          &      4.5   \\
     7.0      &     0.19    &       520      &     1.3          &      3.1   \\
\tableline
    (ML)      &             &                &                  &            \\
\tableline
     2.0      &     0.80    &       460      &     2.3          &      6.0   \\
     0.6      &     1.0     &       420      &     2.9          &      7.0   \\
     0.8      &     1.0     &       420      &     3.9          &      9.2   \\
     1.0      &     1.0     &       420      &     4.7          &     10.4   \\
     1.2      &     1.0     &       420      &     6.0          &     13.0   \\
     1.6      &     1.0     &       420      &     7.8          &     18.0   \\
\end{tabular}

\end{table}



\begin{figure}
\caption{Low temperature photoluminescence (dashed lines) and photoluminescence
excitation spectra (full lines) of the heavy-hole free exciton in three 
InGaAs/GaAs samples with different well width and indium molar fraction. 
The Stokes shift increases from the top to the bottom of the figure, as well 
as the HHFE recombination and absorption linewidths. For ease of comparison, 
a common zero of the energy has been taken at the PLE peak of HHFE 
(at 1.470, 1.400, and 1.471 eV, from top to bottom of the figure).}
\label{Figure1}
\end{figure}

\begin{figure}
\caption{The Stokes shift as a function of the FWHM of the HHFE band measured
in PLE for all the investigated samples. The full line, 
$SS$ = - 0.6 + 0.48\ $W$, has been obtained by a best fit to the data for 
$W \ge$ 6 meV. The dashed line extrapolates the best fit to $W <$ 6 meV. 
Typical error bars are given for the sample with $W$ = 10.4 meV.}     
\label{Figure2}
\end{figure}

\begin{figure}
\caption{The product of the thermal carrier energy k$_BT_c$ and the Stokes 
shift vs $W^2$ for FWHMs of the HHFE band measured in PLE smaller 
than 6 meV (data reported already in Fig. 1). The full line, 
$SS\dot k_BT_c = 0.18\ W^2$, gives the behavior predicted by the 
quasiequilibrium model of Ref. 6. In the inset, the comparison between the
data and the same model is extended to the full set of samples.}
\label{Figure3}
\end{figure}

\begin{figure}
\caption{(a) The Stokes shift as a function of lattice temperature, $T$, for 
the InAs/GaAs sample with $L$ = 1.2 ML. (b) Ratio between the deviation of the
experimental value of $SS$ from that predicted by the quasiequilibrium model
and $SS$ itself as a function of k$_BT_c/W$. $T_c$ is the carrier temperature 
as determined by an exponential fit to the high energy tail of the 
recombination band. The dashed lines are guides to the eye.} 
\label{Figure4}
\end{figure}


\begin{references}
\bibitem{nota1} 
    This equivalence is not strictly true because the PLE lineshape is also
    influenced by carrier dynamics. However, the distortion in the FWHM 
    introduced by this effect can be assumed to be within the present 
    experimental uncertainty in $W$.
\end{references}
\end{document}